\documentstyle[preprint,aps,epsf]{revtex}% Use this for style revtex

\begin{document}
\def\be{\begin{equation}}
\def\ee{\end{equation}}
\def\bea{\begin{eqnarray}}
\def\eea{\end{eqnarray}}
\def\oupb{UPB\ }
\def\pb{PB\ }
\def\eps{\epsilon}
\newcommand{\ket}[1]{| #1 \rangle}
\newcommand{\bra}[1]{\langle #1 |}
\newcommand{\braket}[2]{\langle #1 | #2 \rangle}
\newcommand{\proj}[1]{| #1\rangle\!\langle #1 |}
\newcommand{\ba}{\begin{array}}
\newcommand{\ea}{\end{array}}
\newtheorem{theo}{Theorem}
\newtheorem{defi}{Definition}
\newtheorem{lem}{Lemma}
\newtheorem{exam}{Example}
\newtheorem{prop}{Property}

\author{Barbara M. Terhal}

\title{Bell Inequalities and The Separability Criterion}

 \address{\vspace*{1.2ex}
            \hspace*{0.5ex}{IBM Research Division, Yorktown Heights, NY 10598, US}\\
Email: {\tt terhal@watson.ibm.com}} 

\date{\today}

\maketitle
\begin{abstract}
We analyze and compare the mathematical formulations of the criterion for 
separability for bipartite density matrices and the Bell inequalities. We show 
that a violation of a Bell inequality can formally be expressed as a 
witness for entanglement. We also show how the criterion for separability and the existence of 
a description of the state by a local hidden variable theory, become 
equivalent when we restrict the set of local hidden variable theories to 
the domain of quantum mechanics. This analysis sheds light on the 
essential difference between the two criteria and may help us in understanding whether there exist entangled states for which the statistics of the outcomes of all possible local measurements can be described by a local hidden variable theory.
\end{abstract}

\vspace{1cm}
\noindent PACS: 03.65.Bz, 03.67.-a \\
\noindent Keywords: Bell Inequalities, Quantum Entanglement, Quantum Information Theory

\section{Introduction}

The advent of quantum information theory has initialized the development 
of a theory of bipartite and multipartite entanglement. Pure state 
entanglement has been recognized as an essential resource in performing tasks 
such as teleportation \cite{tele}, distributed quantum computing or in 
solving a classical communication problem \cite{bcw98}. 

Central in the theory of bipartite mixed state entanglement is 
the convertibility of bipartite mixed state entanglement 
to pure state entanglement by local operations and classical communication.
This involves a question of both a qualitative form --can any entanglement be 
distilled from a quantum state? -- as well as of a quantitative form, --how much pure state entanglement can we distill out of a quantum state?
It has been found that there exist entangled quantum states from which no 
pure state entanglement can be distilled. The first steps have been made to 
classify these so called bound entangled states of which 
there are two kinds, bound entanglement with positive partial transposition (PPT) \cite{bepawel,upb1,upb2} and negative partial transposition \cite{nptnond1,nptnond2}. 
In these studies the central underlying motivation is the use of 
entanglement in computational and information processing tasks. 

The question of whether quantum mechanics provides a complete description
of reality underlies the formulation of Bell's original inequality 
\cite{bell}. In 1964 Bell formulated an inequality which 
any local hidden variable theory obeys. He showed however that the EPR (Einstein-Podolsky-Rosen) singlet state 
\be
\ket{\Psi^-}=\frac{1}{\sqrt{2}}\left(\ket{01}-\ket{10}\right)
\label{eprsing}
\end{equation}
would violate the inequality. Even though local 
hidden variable theories were not formulated for this purpose, it 
can easily been understood that the question of whether an entangled state is 
a useful resource in quantum information processing is related 
to the question of whether there exists a local hidden variable model 
for the entangled state \cite{bct:simulating}. This is of importance in the use of entanglement 
in classical communication protocols. Assume that 
Alice and Bob possess a set of entangled states. Assume as well that 
Alice and Bob can only communicate classically with each other. If there 
would exist a local hidden variable model describing the outcomes 
for {\em any} local measurement or sequence of local measurements that 
Alice and Bob could perform on these states, then
the sharing of this set of entangled states could never give rise 
to a communication protocol that is more efficient than any 
classical communication protocol in which Alice and Bob share 
an unlimited amount of random bits. The idea is of course that Alice and 
Bob, instead of using the entangled states, could replace these by 
the local hidden variable model to carry out their protocol. 

In this paper we explore the similarity and the essential difference 
between Bell inequalities and the separability criterion (the necessary 
and sufficient criterion of a bipartite density matrix to be separable).
We first show how to map the violation of a general Bell inequality onto an 
observable ${\rm H}$ that functions as a witness for the entanglement of the state.
Then we show that, if we restrict the set of local hidden variable 
theories to ones that are consistent with quantum mechanics, the two 
criteria are equivalent; that is, a state violates a Bell 
inequality of this restricted kind if and only if the state is 
entangled.

\section{The Separability Criterion}

The set $B({\cal H}_A \otimes {\cal H}_B)$ denotes the set of operators on 
a Hilbert space ${\cal H}_A \otimes {\cal H}_B$. The set $B({\cal H}_A \otimes {\cal H}_B)^+$ denotes the subset of positive semidefinite operators, i.e.
 the unnormalized density matrices on ${\cal H}_A \otimes {\cal H}_B$.
A bipartite density matrix $\rho \in B({\cal H}_A \otimes {\cal H}_B)^+$ 
is called separable if and only if there exists a decomposition of $\rho$ 
into an ensemble of product states, i.e. 
\be
\rho=\sum_i p_i \ket{\phi_i^A}\bra{\phi_i^A} \otimes  \ket{\phi_i^B}\bra{\phi_i^B}, 
\ee
where $p_i \geq 0$. The Horodecki family has formulated the necessary and sufficient conditions 
for separability of a bipartite density matrix: 

\begin{lem}\cite{nec_horo}
A bipartite density matrix $\rho \in B({\cal H}_A \otimes {\cal H}_B)^+$ is entangled if and only if there exists a Hermitian operator ${\rm H} \in B({\cal H}_A \otimes {\cal H}_B)$ with the properties:
\be
\ba{lr}
{\rm Tr}\,{\rm H}\,\rho < 0 & {\rm and} \;\;{\rm Tr}\,{\rm H}\,\sigma \geq  0, 
\ea
\end{equation}
for all separable density matrices $\sigma \in B({\cal H}_A \otimes {\cal H}_B)^+$.
\label{lemhor}
\end{lem}

The lemma follows from basic theorems in convex analysis \cite{rocka}. 
The proof invokes the existence of a separating hyperplane 
between the compact convex set of separable density matrices on 
${\cal H}_A \otimes {\cal H}_B$ and a point, the entangled density matrix $\rho$, that does not belong to it. This separating hyperplane is characterized
by the vector ${\rm H}$ that is normal to it; the hyperplane is the set of 
density matrices $\tau$ such that ${\rm Tr}\,{\rm H}\,\tau=0$. 

From a physics point of view, the Hermitian operator 
${\rm H}$ is the observable that would reveal the entanglement of a density 
matrix $\rho$. We will call ${\rm H}$ an entanglement witness. 
The lemma tells us that there exists such an observable ${\rm H}$ for any 
entangled bipartite density matrix \footnote{Even though this witness 
will exist when $\rho$ is entangled, deciding whether $\rho$ is entangled
by performing sequences of measurements on $\rho$ can be a formidable task. In ${\cal H}_2 \otimes {\cal H}_2$ 
and ${\cal H}_2 \otimes {\cal H}_3$ it is sufficient to consider witnesses of the form 
${\rm H}=({\bf 1} \otimes T)(\ket{\psi} \bra{\psi})$ where $T$ is matrix transposition and $\ket{\psi}$ is some entangled state which will depend on $\rho$. 
For higher dimensions, say ${\cal H}_N \otimes {\cal H}_N$, it is sufficient to
consider witnesses of the form ${\rm H}=({\bf 1} \otimes {\cal P})(\ket{\psi} \bra{\psi})$ where 
${\cal P}$ is some extremal positive map which is not completely positive.
We do however not have a classification of the set of (extremal) positive maps for dimensions other than ${\cal H}_2 \otimes {\cal H}_3$ and ${\cal H}_2 \otimes {\cal H}_2$.}.
We remark that the choice of `origin' of the separation is irrelevant here;
the pair of inequalities ${\rm Tr}\, {\rm H}\, \rho < c$ and ${\rm Tr}\, {\rm H}\, \sigma \geq c$ for all separable $\sigma$, for $c$ other than 0, can be obtained 
by adding a term $c {\bf 1} \otimes {\bf 1}$ to an entanglement witness 
${\rm H}$. 

\section{Bell Inequalities}

We now turn to the formulation of the Bell inequalities. The question of whether 
quantum mechanics provides a complete description
of reality underlies the formulation of Bell's original inequality 
\cite{bell}. The issue is whether the results of measurements 
can be described by assuming the existence of a classical local 
hidden variable. The variable is hidden as its value 
cannot necessarily be measured directly; the average outcome of any measurement is 
a statistical average over different values that this hidden variable can take. The locality of this variable is required by the locality of classical
physics \footnote{No information can travel faster than the speed of light.}. Bell demonstrated that for the state in Eq. (\ref{eprsing}), the EPR singlet 
state, there exists a set of local measurements performed by two parties, Alice and Bob, whose outcomes {\em cannot} be described by any local hidden variable theory. The first experimental verification of his result with 
independently chosen measurements for Alice and Bob was carried out by Aspect {\em et al.} \cite{aspect}. Since Bell's result, much attention has 
been devoted to finding stronger ``Bell inequalities '', that is, inequalities 
that demonstrate the nonlocal character of other entangled states, pure 
{\em and} mixed. It has been found that any bipartite pure entangled state 
violates some Bell inequality \cite{peresbook}. The situation for mixed states is less clear. Whether or not a state violates a Bell inequality may depend 
on what local operations and measurements one allows Alice and Bob to perform. 
For example, it was shown by Werner \cite{werner:lhv} that a 
class of bipartite entangled states in ${\cal H}_d \otimes {\cal H}_d$ could 
be described by a local hidden variable theory when Alice and Bob were to  
perform all possible local von Neumann measurements. Popescu \cite{popescu:nolhv} showed however that with a sequence of measurements on the state a 
violation of a Bell inequality could be found. At its 
most general, we have to assume that any set of local operations and 
classical communication can be performed by Alice and Bob prior to their 
testing of a violation of a Bell inequality.
Then it follows that any set of bipartite mixed states that 
can be distilled \cite{bdsw} will violate a Bell inequality. The distillability 
makes it possible to map the set of mixed entangled states onto a (smaller) set of pure entangled states after which a pure-state Bell-inequality test will reveal their nonlocal character. It is an open question of whether a violation of a Bell inequality can be found for the bipartite or multipartite {\em bound} entangled states (which by definition are undistillable); results in Ref. \cite{werner:pptlhv} (see also Ref. \cite{peres:bell}) indicate that it might be hard (and perhaps impossible) to find a 
violation for bound entangled states with positive partial transpose (PPT).

In what follows we consider the formulation of Bell inequalities when 
two parties, Alice and Bob, perform a set of measurements on a given 
bipartite quantum state $\rho$; we will not consider sequences of measurements.

Interestingly, this general formulation of Bell inequalities
\cite{peres:bell,garg,pitowsky} has great similarity with the separability 
criterion of Lemma \ref{lemhor} and there exists a relation 
between the two.
The general formulation of Bell inequalities comes about in the following way.
We will consider only bipartite states here, but the formulation also 
holds for multipartite states.
Let ${\cal M}_1^A,\ldots {\cal M}_{n_A}^A$ be a set of possible measurements for Alice and 
${\cal M}_1^B,\ldots {\cal M}_{n_B}^B$ be a set of measurements for Bob. 

Each measurement is characterized by its Positive Operators Elements corresponding to the possible outcomes. We write the Positive Operators Elements for the 
$i$th Alice measurement with $k$ outcomes as 
\bea
{\cal M}_i^A=(E_{i,1}^A,E_{i,2}^A,\ldots, E_{i,{k(i)}}^A), & \sum_{m=1}^{k(i)} E_{i,m}^A={\bf 1}, & E_{i,m}^A \geq 0,
\end{eqnarray}
and similarly for the $j$th measurement of Bob,
\bea
{\cal M}_j^B=(E_{j,1}^B,E_{j,2}^B,\ldots, E_{j,l(j)}^B), & \sum_{m=1}^{l(j)} E_{j,m}^B={\bf 1}, & E_{j,m}^B \geq 0.
\end{eqnarray}
Let $\vec{P}$ be a vector of probabilities of outcomes of measurements by 
Alice and Bob on a quantum state $\rho$. The vector $\vec{P}$ has three 
parts denoted with the components $(P_{A:i|k,B:j|l},P_{A:i|k},P_{B:j|l})$. 
For example, when Alice has 
two possible measurements with two outcomes each and Bob has one measurement with 
three outcomes, $\vec{P}$, according to quantum mechanics, will be a 12+4+3 component vector with its
components equal to
\be
\ba{l}
P_{A:i|k,B:j|l}={\rm Tr}\, E_{i,k}^A \otimes E_{j,l}^B \,\rho, \\
P_{A:i|k}={\rm Tr}\, E_{i,k}^A \otimes {\bf 1} \,\rho, \\
P_{B:j|l}={\rm Tr}\, {\bf 1} \otimes E_{j,l}^B \,\rho,
\ea
\end{equation}
for $i=1,2$, $k=1,2$ and $j=1$, $l=1,2,3$. We call $\vec{P}$ the event vector. 

Let $\lambda$ be a local hidden variable. We choose $\lambda$ such that when $\lambda$ takes a specific value, each
measurement outcome is made either impossible or made to occur with 
probability 1. In other words, given a value of $\lambda$ a probability of 
either 0 or 1 is assigned to Alice's outcomes and similarly for Bob.
Then we choose $\lambda$ to take as many values as are needed to produce 
all possible patterns of 0s and 1s, all Boolean vectors. These outcome 
patterns are denoted as Boolean vectors $\vec{B}_{\lambda}^A$ and 
$\vec{B}_{\lambda}^B$. For example, when Alice has three measurements each 
with two outcomes there will be $2^6$ vectors $\vec{B}_{\lambda}^A \in \{0,1\}^6$. The vector $\vec{B}_{\lambda}^A$ has of course the same number of entries 
as Alice's part of the event vector $\vec{P}_A$ and similarly for Bob.
The locality constraint comes in by requiring that the joint vector $\vec{B}_{\lambda}^{AB}$ is a product vector, i.e. 
$\vec{B}_{\lambda}^{AB}=\vec{B}_{\lambda}^A \otimes 
\vec{B}_{\lambda}^B$. The total vector is denoted as $\vec{B}_{\lambda}=
(\vec{B}_{\lambda}^{AB},\vec{B}_{\lambda}^{A},\vec{B}_{\lambda}^{B})$.
Our example will serve to elucidate the idea. When, as before, Alice has 
two measurements each with two outcomes and Bob has one measurement with three outcomes, there exists a $\lambda$ such that such that the vector $\vec{B}_{\lambda}$ is
\be
\vec{B}_{\lambda}=[(\overbrace{1,0,}^{{\cal M}_1^A}\overbrace{0,1}^{{\cal M}_2^A}) \otimes 
 (\overbrace{0,1,0}^{{\cal M}_1^B}), (1,0,0,1),(0,1,0)].
\end{equation}
We denote the vector $\vec{B}_{\lambda=\lambda_1}$, when $\lambda$ takes the value 
$\lambda_1$ as $\vec{B}_{\lambda_1}$. Any local hidden variable theory 
can be represented as a vector $\vec{V}$:
\be
\vec{V}=\sum_{i} p_i\, \left(\vec{P}^A_i \otimes \vec{P}^B_i,\vec{P}^A_i, \vec{P}^B_i\right),
\end{equation}
with $p_i \geq 0$ and $\vec{P}^A_i$ and $\vec{P}^B_i$ are vectors of (positive) numbers. These vectors $\vec{V}$ are convex combinations of the vectors 
$\vec{B}_{\lambda_1}, \ldots, \vec{B}_{\lambda_N}$, where $N$ is such that 
$\vec{B}_{\lambda}^A$ and $\vec{B}_{\lambda}^B$ are all possible Boolean vectors (see Ref. \cite{pitowsky}):
\be
\vec{V}=\sum_i q_i \vec{B}_{\lambda_i},
\end{equation}
with $q_i \geq 0$. Note that we have not constrained the vectors $\vec{B}_{\lambda}^A$ or $\vec{B}_{\lambda}^B$ such that the total probability for each measurement is $1$. Also note that at the same time we include the marginal 
`probabilities' $\vec{P}^A_i$ and $\vec{P}^B_i$ in the total vector $\vec{V}$. 
The alternative, but equivalent formulation, is the one (see for example Ref. \cite{peres:bell}) in which we only consider convex combinations of the joint 
probabilities $\vec{P}^A_i \otimes \vec{P}^B_i$ from which the marginals 
follow directly, and we restrict the Boolean vectors $\vec{B}_{\lambda}^A$ 
etc. to be such that the sum of the probabilities for each measurement is 
equal to 1. 

Thus we see that the set of local hidden variable theories forms a convex cone $L_{LHV({\cal M})}$. 
The label ${\cal M}$ is a reminder that the cone depends on the chosen measurements for Alice or Bob, in particular the number of them 
and the number of outcomes for each of them. The vectors $\vec{B}_{\lambda_i}$ 
are the extremal rays \cite{peres:bell} of $L_{LHV({\cal M})}$.
The question then of whether the probabilities of the outcomes of the chosen set of measurements on a density matrix $\rho$ can be reproduced by a local hidden variable theory, is equivalent to the question whether or not
\be
\vec{P} \in L_{LHV({\cal M})}.
\label{pinx}
\end{equation}
It is not hard to see that all separable {\em pure} states have event vectors $\vec{P} \in L_{LHV({\cal M})}$ as the event vector $\vec{P}$ for a separable pure state has a product structure $\vec{P}=(\vec{P}_A
 \otimes \vec{P}_B,\vec{P}_A,\vec{P}_B)$. It follows that all separable states have event vectors in 
$L_{LHV({\cal M})}$, as they are convex combinations of separable pure states.
What about the entangled states? We can use the Minkowski-Farkas lemma for convex sets in ${\bf R}^n$ \cite{rocka}. 
The lemma implies that $\vec{P} \notin L_{LHV({\cal M})}$ if and only if there exists a vector $\vec{F}$ such that
\be
\ba{lr}
\vec{F} \cdot \vec{P} < 0 & {\rm and} \;\;\forall\;\lambda_i\; \left[\vec{F} \cdot \vec{B}_{\lambda_i} \geq  0 \right]. 
\ea
\label{farkas}
\end{equation}
The equation $\forall\;\lambda_i\; \left[\vec{F} \cdot \vec{B}_{\lambda_i} \geq  0\right]$ 
is a Bell inequality. The equation $\vec{F} \cdot \vec{P} < 0$ corresponds to
the violation of a Bell inequality. Thus, finding a set of measurements and 
exhibiting the vector $\vec{F}$ with the properties of Eq. (\ref{farkas}) is equivalent to finding a violation of a Bell inequality. If one can prove that for a density matrix $\rho$ no such sets of inequalities of the form Eq. (\ref{farkas}) for all possible measurement schemes can be found, then it follows 
that $\rho$ can be described by a local hidden variable theory. 

There is a nice correspondence between Eq. (\ref{farkas}) and 
Lemma \ref{lemhor}, captured in the following construction: Given a 
(Farkas) vector $\vec{F}$ of Eq. (\ref{farkas}) and a set of measurements 
${\cal M}$ for a bipartite entangled state $\rho$, one can construct an entanglement witness for $\rho$ as in Lemma \ref{lemhor}. Denote the components of the Farkas vector $\vec{F}$ as $(F_{A:i|k,B:j|l},F_{A:i|k},F_{B:j|l})$. 
Then 
\be
{\rm H}=\sum_{i,k,j,l} F_{A:i|k,B:j|l} E_{i,k}^A \otimes E_{j,l}^B
+\sum_{i,k} F_{A:i|k} E_{i,k}^A \otimes {\bf 1} +
\sum_{j,l} F_{B:j|l} {\bf 1} \otimes E_{j,l}^B
,
\label{makeh}
\end{equation}
where $E_{i,k}^A$ is the positive operator of the $i$th measurement with outcome $k$ for Alice and similarly for Bob. With this construction $\vec{F} \cdot \vec{P}={\rm Tr}\,{\rm H}\,\rho$. Also, one has ${\rm Tr}\,{\rm H}\,\sigma \geq 0$ for any separable density matrix 
$\sigma$ as $\vec{P}_{\sigma} \in L_{LHV({\cal M})}$ for all separable density matrices $\sigma$. 
Thus a violation of a Bell inequality for a bipartite density matrix $\rho$ 
can be reformulated as an entanglement witness ${\rm H}$ for $\rho$. 

We can illustrate the construction with the well known Bell-CHSH 
inequality \cite{chsh} for two qubits. It is convenient to start with the following form 
\be
\vec{F} \cdot \vec{P}= p_a-p_{ab}+p_{b'}-p_{a'b'}+p_{a'b}-p_{ab'} \geq 0,
\ee
where $a$ and $a'$ characterize Alice's two measurements: i.e. Alice 
measures the eigenvalues, $+1$ or $-1$, of the observables $\vec{a} \cdot \vec{\sigma}$ and  $\vec{a'} \cdot \vec{\sigma}$ where $\vec{a}$ and $\vec{a'}$ are unit vectors. Similarly, Bob can measure the eigenvalues of $\vec{b} \cdot \vec{\sigma}$ and  $\vec{b'} \cdot \vec{\sigma}$. The probability $p_{ab}$ 
corresponds to the probability that both Alice and Bob find outcome $+1$ 
for observable $\vec{a} \cdot \sigma$ and $\vec{b} \cdot \sigma$. Similarly 
do they other probabilities correspond to finding the $+1$ eigenvalue of the 
other measurements.
Now it follows with Eq. (\ref{makeh}) that ${\rm H}$ is equal to 
\bea
{\rm H}=\frac{1}{2}[{\bf 1}+\vec{a} \cdot \vec{\sigma}] \otimes {\bf 1} -
\frac{1}{2}[{\bf 1}+\vec{a} \cdot \vec{\sigma}] \otimes \frac{1}{2}[2 {\bf 1}+(\vec{b}+\vec{b'})\cdot \vec{\sigma}]
\nonumber \\
+{\bf 1} \otimes \frac{1}{2}[{\bf 1}+\vec{b'} \cdot \vec{\sigma}]-\frac{1}{2}[{\bf 1}+\vec{a'} \cdot \vec{\sigma}] \otimes \frac{1}{2}(\vec{b'}-\vec{b})\cdot \vec{\sigma},
\eea
which can be rewritten as 
\be
{\rm H}=\left(2{\bf 1}-\vec{a} \cdot \vec{\sigma}\otimes (\vec{b}+\vec{b'})\cdot \vec{\sigma}-\vec{a'} \cdot \vec{\sigma}\otimes (\vec{b'}-\vec{b})\cdot \vec{\sigma}\right)/4.
\ee
Here we recognize the Bell operator ${\cal B}=\vec{a} \cdot \vec{\sigma}\otimes (\vec{b}+\vec{b'})\cdot \vec{\sigma}-\vec{a'} \cdot \vec{\sigma}\otimes (\vec{b'}-\vec{b})\cdot \vec{\sigma}$, which was first introduced in Ref. \cite{braunstein_etal:bell}. 

The relation between a Bell inequality and the separability condition 
gives a clue about what to look for when trying to find a violation of 
a Bell inequality for bound entangled PPT states. Every Hermitian operator 
${\rm H}$ which has the property that 
\be
{\rm Tr} \, {\rm H}\, \sigma \geq 0,
\ee
for all separable states $\sigma$ can be written as 
\be
{\rm H}=({\bf 1} \otimes {\cal P}) (\ket{\Psi} \bra{\Psi}),  
\ee
where $\ket{\Psi}$ is a maximally entangled state (see Ref. \cite{nptnond1}) 
and ${\cal P}$ is a positive map. Since ${\rm Tr} {\rm H}\, \rho < 0$ and $\rho$ is a PPT state, it follows
that the positive map ${\cal P}$ {\em cannot} be related to the transposition map $T$ in the following way
\be
{\cal P} \neq {\cal S}_1 +{\cal S}_2 \circ T, 
\ee 
where ${\cal S}_1$ and ${\cal S}_2$ are completely positive maps. In other words, ${\cal P}$ is not a decomposable positive map. We note that for bound 
entangled states which are based on unextendible product bases, the 
entanglement witness ${\rm H}$ is known \cite{terhalposmap}. It is 
possible to try to search numerically for violations of Bell inequalities 
for the corresponding bound entangled states by decomposing such a
witness into positive operators of POVM measurements and coefficients of a 
Farkas vector.

\section{Restricted Local Hidden Variables}

One may now ask the following question:
Given an entanglement witness ${\rm H}$ for a bipartite density matrix $\rho$, does 
there exist a decomposition of ${\rm H}$ into a set of measurements and a vector 
$\vec{F}$ as in Eq. (\ref{makeh}), that leads to a violation of a Bell 
inequality for $\rho$? The reason for the discrepancy 
between the inequalities of Lemma \ref{lemhor} and Eq. (\ref{farkas}) is 
that the hidden variable cone $L_{LHV({\cal M})}$ contains more than 
just the separable states; it can also contain vectors which do 
{\em not} correspond to probabilities of outcomes of measurements on a quantum mechanical system. 
If quantum mechanics is correct then we will never find these sets of outcomes. An example of such an unphysical vector is the following. 
Let Alice perform two possible measurements on a 
two-dimensional system. Her first measurement ${\cal M}_1^A$ is a projection 
in the $\{\ket{0}, \ket{1}\}$ basis and her second measurement ${\cal M}_2^A$ is a projection in the $\{\frac{1}{\sqrt{2}}(\ket{0}+\ket{1}),\frac{1}{\sqrt{2}}(\ket{0}-\ket{1})\}$ basis. The hidden variable cone $L_{LHV({\cal M})}$ will contain vectors such as
\be
\vec{B}_{\lambda}=[(\overbrace{1,0,}^{{\cal M}_1^A}\overbrace{0,1}^{{\cal M}_2^A}) \otimes (\ldots), (1,0,0,1),(\ldots)].
\end{equation}
This vector $\vec{B}_{\lambda}$ which assigns a probability 1 to outcome $\ket{0}$ {\em and}
a probability 1 to outcome $\frac{1}{\sqrt{2}}(\ket{0}-\ket{1})$ cannot describe the outcome of these measurements on any quantum mechanical state $\rho$.
%[sketchy formulation] Another situation in which the Boolean vector $\vec{B}_{%\lambda}$ do not represent the probabilities of outcomes of measurements of 
%a quantum system is the following. Alice chooses to perform measurements that %are related to the Kochen-Specker theorem, for example on a three dimensional %system. It is impossible to make a pattern of 0s and 1s in a way that is 
%consistent with the outcomes of measurements on a quantum mechanical system. 

These unphysical vectors play an important role in the 
construction of hidden variable theories for entangled states: their importance is emphasized by the following observation. If we restrict the cone $L_{LHV({\cal M})}$ to contain only vectors that 
are consistent with quantum mechanics, then we can prove that there exists 
a ``violation of a Bell inequality'' for {\em any} entangled state. By this 
we mean the following: We demand that {\em all} 
vectors in the set $L_{LHV({\cal M})}$ correspond to sets of outcomes that 
can be obtained by measurements on a quantum mechanical system in ${\cal H}_A \otimes {\cal H}_B$. Here ${\cal H}_A \otimes {\cal H}_B$ is the Hilbert space 
on which the density matrix, that we attempt to describe with a 
restricted local hidden variable theory, is defined.
We can call this restricted local hidden variable theory a local quantum 
mechanical hidden variable theory. One can prove that in this restricted 
scenario, there will always be a set of measurements under which $\rho$ reveals its nonlocality and its entanglement:

\begin{theo}
Let $\rho$ be a bipartite density matrix on ${\cal H}_A \otimes {\cal H}_B$.
The density matrix $\rho$ is separable if and only if there exists a restricted local hidden variable theory of $\rho$.
\label{theobell}
\end{theo}

{\em Proof} The idea of the proof is the following. All vectors in the 
restricted local hidden variable theory now correspond to outcomes of measurements on a quantum mechanical system. We chose a set of measurements that completely determines a quantum state in a given Hilbert space. Then there is a 
one-to-one correspondence between vectors of measurement outcomes and quantum states. 
Then we show that all vectors in the restricted local hidden variable set 
correspond to measurement outcomes of separable states. Therefore measurement 
outcomes from entangled states do not lie in the set described by a restricted 
local hidden variable theory.

We write the density matrix $\rho$ as
\be
\rho=\sum_{i,j} \mu_{ij} \sigma_i \otimes \tau_j+\sum_{i} \mu_{i}^A \sigma_i \otimes {\bf 1}+\sum_{j} \mu_{j}^B {\bf 1} \otimes \tau_j+c {\bf 1}\otimes {\bf 1},
\end{equation}
where the Hermitian matrices $\{\sigma_i \otimes \tau_j\}_{i=1,j=1}^{d_A^2-1,d_B^2-1},\{\sigma_i \otimes {\bf 1}\}_{i=1}^{d_A^2-1},\{{\bf 1} \otimes \tau_j\}_{j=1}^{d_B^2-1}$, ${\bf 1} \otimes {\bf 1}$ with $d_A=\dim {\cal H}_A$ etc., form a basis for the Hermitian operators on ${\cal H}_A \otimes {\cal H}_B$. 
The constant $c$ is fixed by ${\rm Tr} \rho =1$ and will depend on the other 
parameters $\mu_{ij}$ etc. when the matrices $\sigma_i,\tau_j$ are not traceless. Let $\ket{w^A_{i,k}}$ be the eigenvectors of the matrix $\sigma_i$ 
and $\ket{w^B_{j,l}}$ be the eigenvectors of $\tau_j$. The projector 
onto the state $\ket{w^A_{i,k}}$ is denoted as 
$\pi_{w^A_{i,k}}$ and similarly, the projector onto the state 
$\ket{w^B_{j,l}}$ is denoted as $\pi_{w^B_{j,l}}$. 

Alice and Bob choose a set of measurements such that the probabilities of 
outcomes of these measurements are, according to quantum mechanics, given by
\be
\ba{l}
{\rm Tr}\, \pi_{w^A_{i,k}} \otimes \pi_{w^B_{j,l}}  \,\rho=p_{i,k,j,l}, \\
{\rm Tr}\, \pi_{w^A_{i,k}} \otimes {\bf 1}  \,\rho=p_{i,k}^A, \\
{\rm Tr}\, {\bf 1} \otimes \pi_{w^B_{j,l}}  \,\rho=p_{j,l}^B\;,
\label{probout}
\ea
\end{equation}
for all $i$, $k$, $j$ and $l$. What is important is that they, if they would carry out these
 measurements repeatedly on $\rho$ (a single measurement on each copy of $\rho$), would be able to determine the probabilities $(p_{i,k,j,l},p_{i,k}^A,p_{j,l}^B)$. Then they can uniquely infer from these probabilities the state $\rho$.
We call this set of measurements ${\cal M}_{c}$, a complete set of measurements. Let $L_{LHV({\cal M}_{c})}^r$ be the convex set 
of restricted local hidden variable theories \footnote{Note that $L_{LHV({\cal M}_{c})}^r$ is a set and not a cone, as $\vec{V} \in L_{LHV({\cal M}_{c})}^r$ does not 
imply that $\lambda \vec{V} \in L_{LHV({\cal M}_{c})}^r$ with $\lambda > 0$, 
as we now require that all vectors in $\vec{V}$ correspond to probabilities of 
outcomes of measurements on a quantum mechanical system.}. 
We first consider which density matrices $\rho$ can be described 
by restricted local hidden variable vectors of the form $(\vec{P}_A \otimes \vec{P}_B,\vec{P}_A,\vec{P}_B)$,
where $\vec{P}_A$ ($\vec{P}_B$) is a vector of probabilities $p_{i,k}^A$ ($p_{j,l}^B$). The density matrix $\rho=\rho_A \otimes \rho_B$ 
where $\rho_A={\rm Tr}_B \rho$ and $\rho_B={\rm Tr}_A \rho$ is a solution of the equations
\be
\ba{l}
{\rm Tr}\, \pi_{w^A_{i,k}} \otimes \pi_{w^B_{j,l}}  \,\rho=p_{i,k}^A\, p_{j,l}^B , \\
{\rm Tr}\, \pi_{w^A_{i,k}} \otimes {\bf 1}  \,\rho=p_{i,k}^A, \\
{\rm Tr}\, {\bf 1} \otimes \pi_{w^B_{j,l}}  \,\rho=p_{j,l}^B\;,
\ea
\label{solveeq}
\end{equation}
for all $i,k,j$ and $l$, since
\be
{\rm Tr}\, \pi_{w^A_{i,k}} \otimes \pi_{w^B_{j,l}}  \,\rho=
{\rm Tr}\, \pi_{w^A_{i,k}} \otimes \pi_{w^B_{j,l}} (\rho_A \otimes \rho_B),
\label{sepbell}
\end{equation}
As the set of measurements completely 
determines the density matrix $\rho$ it follows that the solution $\rho=\rho_A \otimes 
\rho_B$ is the only solution of Eq. (\ref{solveeq}) for all $i$, $k$, $j$ and $l$ . Therefore all the restricted local variable vectors of the form $(\vec{P}_A \otimes \vec{P}_B,\vec{P}_A,\vec{P}_B)$ correspond to product states. If follows that 
any convex combination of the restricted local hidden variable vectors $\vec{V}=\sum_{i} p_i \,(\vec{P}_A^i \otimes \vec{P}_B^i,\vec{P}_A^i,\vec{P}_B^i)$ corresponds to a separable state. As the map from the vectors $\vec{P}$ to states 
$\rho$ is one-to-one, this is the only density matrix that corresponds to $\vec{V}$. 
Thus we can conclude that no vector in the convex set $L_{LHV({\cal M}_{c})}^r$ corresponds to an entangled state. On the other 
hand the outcome vector of any separable density matrix lies in $L_{LHV({\cal M}_{c})}^r$ by the argument given below Eq. (\ref{pinx}). This completes the 
proof. $\Box$ 

We are now ready to clarify the relation between the separability criterion, 
Lemma \ref{lemhor}, and Bell inequalities. Theorem \ref{theobell} shows that 
$L_{LHV({\cal M}_{c})}^r$ only contains outcome vectors of 
separable states. We decompose the entanglement witness ${\rm H}$ in terms of 
the vectors $\ket{w^A_{i,k}}$ and $\ket{w^B_{j,l}}$, given by ${\cal M}_{c}$.
\be
{\rm H}=\sum_{i,k,j,l} F_{A:i|k,B:j|l} \pi_{w^A_{i,k}} \otimes  \pi_{w^B_{j,l}}+
\sum_{i,k} F_{A:i|k} \pi_{w^A_{i,k}} \otimes {\bf 1}+
\sum_{j,l} F_{B:j|l} {\bf 1} \otimes  \pi_{w^B_{j,l}}+c {\bf 1}\otimes {\bf 1}.
\end{equation}
This is always possible as the set $\{\sigma_i \otimes \tau_j\}_{i=1,j=1}^{d_A^2-1,d_B^2-1}$,$\{\sigma_i \otimes {\bf 1}\}_{i=1}^{d_A^2-1},\{{\bf 1} \otimes \tau_j\}_{j=1}^{d_B^2-1}$, ${\bf 1} \otimes {\bf 1}$ forms a basis for the Hermitian operators on ${\cal H}_A \otimes {\cal H}_B$. The last term proportional 
to ${\bf 1} \otimes {\bf 1}$ does not play any r\^ole, since it is a 
$\rho$-independent term in the expression ${\rm Tr}\,{\rm H}\, \rho$. Let us 
therefore consider the decomposition of ${\rm H'}={\rm H}-c {\bf 1}\otimes {\bf 1}$.
The coefficients $(F_{A:i|k,B:j|l},F_{A:i|k},F_{B:j|l})$ are real and are identified with the components of the vector $\vec{F}$. We then have an  
equivalence between the separability criterion and a ``violation of a Bell 
inequality'' with restricted local hidden variables:
\be
\ba{rcl}
\vec{F} \cdot \vec{P} & = & {\rm Tr}\,{\rm H'}\,\rho,  \\
& \mbox{and} & \\
\forall\,\vec{V} \in L_{LHV({\cal M}_{c})}^r,\; \vec{F} \cdot \vec{V} \geq  0
& \Leftrightarrow & \forall \mbox{ separable } \sigma,  {\rm Tr}\,{\rm H'}\,\sigma \geq 0. 
\ea
\end{equation}

\section{Conclusion} 
To conclude, we have been able to show that there is an equivalence 
relation between the separability criterion and a weak form of Bell inequality,
namely one that assumes that the variables take a restricted set of values, 
consistent with quantum mechanics. The analysis as 
presented does not resolve the question of whether all entangled states violate 
a Bell inequality in the strong sense, one where the variable can take 
`unphysical' values. In particular, the question whether there can exist 
a violation of a Bell inequality for PPT entangled states remains open.

\section{Acknowledgments}
The author would like to thank Asher Peres and John Smolin for interesting discussions and David DiVincenzo for useful comments and corrections on an earlier draft of this manuscript.

\end{document}